\documentclass[conference,a4paper]{IEEEtran}
\usepackage{cite}

\ifCLASSINFOpdf
   \usepackage[pdftex]{graphicx}
   \graphicspath{{../pdf/}{../jpeg/}}
   \DeclareGraphicsExtensions{.pdf,.jpeg,.png}
\else
   \usepackage[dvips]{graphicx}
   \graphicspath{{../eps/}}
   \DeclareGraphicsExtensions{.eps}
\fi
\usepackage[cmex10]{amsmath}

\usepackage[pass]{geometry}

\usepackage{aastex_hack}

\begin{document}
%
\title{Recent Developments in Measuring Signal and Noise in Phased Array Feeds at {CSIRO}}

\author{\IEEEauthorblockN{
A.~P. Chippendale,\ 
D. McConnell, \ 
K. Bannister, \ 
N. Nikolic, \ 
A.~W. Hotan,\\
K.~W. Smart, \ 
R.~D. Shaw, \ 
D.~B. Hayman, \ 
S.~G. Hay
}                                     
CSIRO, Sydney, Australia, Aaron.Chippendale@csiro.au}



\maketitle

\begin{abstract}
We describe recent developments in measuring both signal and noise in phased array feeds for radio astronomy at CSIRO. We introduce new techniques including aperture array noise measurements with beamforming weights matched to a reflector's focal field.  Weights are calculated via antenna-range and in-reflector measurements.  We also describe the separation of system temperature and aperture efficiency via drift scans.
\end{abstract}

{\smallskip \keywords phased array feed, antenna measurements, radio astronomy, electromagnetic simulation, ASKAP}

%
\IEEEpeerreviewmaketitle

\section{Introduction}
Phased array feeds (PAFs) can significantly increase the survey speed of radio telescopes. A PAF is a dense array of antenna elements at the focus of a concentrator that, with digital beamforming, can produce multiple simultaneous antenna beams of high sensitivity throughout a wide field of view. The first example in service is CSIRO's Australian Square Kilometre Array Pathfinder (ASKAP) Telescope system \cite{Hotan2014,Hampson2012}, which will soon be joined by the Westerbork Synthesis Radio Telescope fitted with PAFs by the APERTIF project \cite{Cappellen2010}.

Measuring the performance of PAFs is crucial to their development and adoption. Survey speed is largely determined by the PAF beam sensitivities $A_\text{e}/T_\text{sys}$   throughout the field of view. However measuring components of the effective area $A_\text{e}$  and system noise temperature $T_\text{sys}$ is important for understanding behaviour and verifying the designs of the reflector and feed components. The noise contribution of the PAF low-noise amplifiers (LNAs) is particularly important and must be minimised when LNA signals are beamformed with weights matched to the focal-field illumination.
	
At EuCAP 2014 a review was presented on the PAF measurement challenges and techniques used at CSIRO \cite{Hayman2014}. Here we describe our more recent developments.

\section{Methods}
\subsection{Drift Scans}
In this method the sensitivities of the beams in a PAF interferometer are first determined from correlations between PAF beams from different reflector antennas. These are measured with both beams centred on an astronomical point source of known flux. Our new measurements include drift scans \cite{McConnell2015} with the antennas held fixed while Earth rotation scans the beams through emission of varying strength near the Galatic plane. At time $i$ the beam sensitivity can be expressed as
\begin{equation}
\label{eq:driftscan}
\left(\frac{A_\text{e}}{T_\text{sys}}\right)_i = \frac{\eta_\text{rad}\eta_\text{ap}A_\text{p}}{\eta_\text{rad}\left(T_{\text{sky,}i}+ T_\text{spill}\right)+\left(1-\eta_\text{rad}\right)T_\text{p} + T_\text{rec}}%
\end{equation}
where $A_\text{p}$ is the known antenna aperture area, $\eta_\text{ap}$ and $\eta_\text{rad}$  are the aperture and radiation efficiencies, $T_\text{p}$ is the physical temperature of the PAF and $T_\text{rec}$ is the receiver temperature. 

The term $T_{\text{sky,}i}+ T_\text{spill}$ is the noise contribution from sources external to the antenna.  The external spillover contribution $T_\text{spill}$  is from sky and ground radiation that enters the PAF via paths other than the focusing of the reflector.  For drift scans, where the antennas are fixed with respect to the ground, we assume that $T_\text{spill}$ is constant as it is dominated by ground radiation at decimetre wavelengths.  However, there will be time variation in $T_\text{spill}$  when the PAF directly glimpses the Galactic plane beyond the edge of the reflector.  

The time-varying external contribution $T_{\text{sky,}i}$ is from sky emission received via the main lobe and near sidelobes of the PAF beam.  We estimate $T_{\text{sky,}i}$ using the Global Sky Model \cite{deOliveira-Costa2008} and a model PAF beam.  The unknown time-invariant parameters $\eta_\text{ap}$ and $\left[T_\text{spill}+\left(1-\eta_\text{rad}\right)T_\text{p} + T_\text{rec}\right]/\eta_\text{rad}$ are estimated by minimising the mean-square error in \eqref{eq:driftscan} across the scan. 

\subsection{Beamforming Conversions}
Previously we have measured the noise temperature of arrays with beamforming for maximum sensitivity for plane-wave illumination of the array \cite{Chippendale2014,Chippendale2015a,Hayman2014}. This is done in aperture-array (AA) test facilities. The array is placed on the ground facing the sky. A signal source is suspended above the array, sufficiently approximating plane-wave illumination. The beamformed noise temperature is deduced using the Y-factor method, using the sky as the cold load and microwave absorber as the hot load. This is convenient compared to installing the array at the focus of a reflector antenna. However the restriction to plane-wave signal illumination is limiting. 

Here we provide conversion factors that can be applied in the on-ground tests to configure the beamformer for maximum sensitivity for focal-field signal illumination of the array, as installed on a reflector antenna. In this approach the noise temperatures are obtained from the beamformed Y-factors
\begin{equation}
  \label{eq:yfact}
  Y = {P_\text{h}}/{P_\text{c}} = {\bar{\mathbf{w}}^t\mathbf{G}_\text{hot}\mathbf{w}}/{\bar{\mathbf{w}}^t\mathbf{G}_\text{cold}\mathbf{w}}
\end{equation}
where $\mathbf{G}_\text{cold}$  and $\mathbf{G}_\text{hot}$ are the array covariance matrices measured with the array viewing the cold sky and the hot absorber respectively.  We have used the beamforming convention of \cite{hay2010} where $\mathbf{w}$ is a column vector of beamforming weights, the beamformed voltage is $v_\text{out}=\mathbf{w}^t\mathbf{v}$, and $\mathbf{v}$ is a column vector of voltages sampled by the array.  The array covariance matrix is defined as $\mathbf{G}=\left<\bar{\mathbf{v}}\mathbf{v}^t\right>$ and the expectation denoted by $\left<\cdot \right>$ is evaluated over a finite integration time in our experiments\footnote{Discrepancies with \cite{Hotan2014,Chippendale2014,Chippendale2015,Chippendale2015a} arise only from the beamforming convention and notation chosen here for the antenna engineering audience.}.

With plane-wave signal illumination of the array from its boresight, the maximum sensitivity AA weights are \cite{Applebaum1976}
\begin{equation}
\mathbf{w}_\text{AA}=\mathbf{G}^{-1}_\text{cold}\bar{\mathbf{s}}_\text{AA}
\end{equation}
where $\bar{\mathbf{s}}_\text{AA}$ is the complex conjugate of the vector of received signals. The weights for maximum sensitivity focal-field signal illumination of the array, as a PAF at the focus of a given reflector, can be obtained via
\begin{equation}
  \label{eq:pafweights}
  \mathbf{w}_\text{PAF}=\mathbf{G}^{-1}_\text{cold}\bar{\mathbf{s}}_\text{PAF}
\end{equation}
where
\begin{equation}
  \label{eq:pafsignal}
  \mathbf{s}_\text{PAF} = \mathbf{C}\mathbf{s}_\text{AA}
\end{equation}
and $\mathbf{C}$ is a diagonal matrix with $j^\text{th}$ diagonal element 
\begin{equation}
  \label{eq:convfactors}
  c_j = {s}_{\text{PAF},j}/{s}_{\text{AA},j}
\end{equation}
where ${s}_{\text{PAF},j}$ is the received signal at array element $j$ with focal-field signal illumination and ${s}_{\text{AA},j}$ is the received signal at array element $j$ with plane-wave signal illumination.  Substituting \eqref{eq:pafweights} and \eqref{eq:pafsignal} into \eqref{eq:yfact} and rearranging gives
\begin{eqnarray}
  Y = \left.{\mathbf{s}^t_\text{AA}\hat{\mathbf{G}}_\text{cold}^{-1}\hat{\mathbf{G}}_\text{hot}\hat{\mathbf{G}}_\text{cold}^{-1}\bar{\mathbf{s}}_\text{AA}}\middle/{\mathbf{s}^t_\text{AA}\hat{\mathbf{G}}_\text{cold}^{-1}\bar{\mathbf{s}}_\text{AA}}\right.  \\
  \hat{\mathbf{G}}_\text{cold}= \bar{\mathbf{C}}^{-1} \mathbf{G}_\text{cold}  {\mathbf{C}}^{-1}\\
  \hat{\mathbf{G}}_\text{hot}= \bar{\mathbf{C}}^{-1} \mathbf{G}_\text{hot}  {\mathbf{C}}^{-1}
\end{eqnarray}
showing that the conversion is equivalent to pre and post-multiplying covariance matrices by $\bar{\mathbf{C}}^{-1}$ and $\mathbf{C}^{-1}$ respectively. The conversion factors \eqref{eq:convfactors} are obtained in two ways:  
\subsubsection{In-reflector measurements with astronomical and on-reflector sources}
\label{sec:in_reflector}
In this method the conversion factors \eqref{eq:convfactors} are obtained from in-reflector measurements. ${s}_{\text{PAF},j}$  is the received signal at array element $j$  with focal-field signal illumination, obtained by observing an astronomical point source. ${s}_{\text{AA},j}$  is the received signal at array element $j$ with plane-wave signal illumination, obtained by transmitting from an antenna located at the vertex of the reflector. This approach has been implemented using a noise source on an ASKAP antenna equipped with a Mark II PAF \cite{Hampson2012, Chippendale2015} for which the AA signal vector and covariance matrices have been obtained \cite{Chippendale2015a}.  The AA covariance matrices are also referred to a plane-wave signal vector to calibrate gain and path-length changes in the backend electronics between AA and in-reflector measurements.
\subsubsection{Radiation-pattern measurements and reflector modelling}
The ratios \eqref{eq:convfactors} are independent of backend signal chain components following the LNAs. Therefore they can be determined from anechoic-chamber radiation-pattern measurements made without the digital backend.  Thus in \eqref{eq:convfactors} ${s}_{\text{AA},j}$   is the measured boresight radiation pattern at LNA port $j$  and ${s}_{\text{PAF},j}$  is the corresponding radiation pattern of the combined array and reflector system. We determine ${s}_{\text{PAF},j}$ using the measured array radiation patterns and analysis of scattering by the reflector. Various reflector parameters can be investigated, including focal-length-to-diameter ratio, single or dual reflectors, offset-fed or shaped reflectors.
	This approach also allows improved estimation of the beamformed radiation pattern and its interaction with the hot load. This becomes more important when the beamformer is configured for focal-field illumination, since the array beam is broader relative to the absorber hot load with PAF rather than AA weights.  This approach has been implemented by measurement of all port radiation patterns of a prototype ASKAP Mk. II PAF.

\section{Weights from in-reflector measurements}
\subsection{Technique}
To calibrate in-reflector PAF weights with respect to an on-reflector calibration source, we first form the diagonal calibration matrix $\mathbf{D}$ whose $j^\text{th}$ diagonal element is 
\begin{equation}
  \label{eq:calfactors}
  d_j = {s}_{\text{cal},j}/{s}_{\text{cal},*}
\end{equation}
where ${s}_{\text{cal},j}$ is a measure of the radiated calibration signal received by PAF port $j$ and ${s}_{\text{cal},*}$ is a measure of the radiated calibration signal received by a reference PAF port.  We chose the reference port to be 141 in our numbering system \cite{Reynolds2014}.  

We use the system described in \cite{Hayman2010} to measure the relative port responses $d_{j}$.  A source at the vertex of the reflector radiates stable broadband noise at a low level into all the PAF elements.  A sample of the calibration signal $v_{cal}$ is then correlated with the PAF port outputs.  This allows relative complex-valued port responses $d_{j}$ to be estimated at each instant by evaluating \eqref{eq:calfactors} in the form of $ d_j = \left< \bar{v}_j v_{cal}\right> / \left< \bar{v}_* v_{cal}\right>$.

Maximum sensitivity PAF weights are calculated following the method in \cite{Chippendale2015} via a measurement of covariance $\mathbf{G}_\text{on}$ towards an astronomical source of known flux and a measurement $\mathbf{G}_\text{off}$ towards nearby empty sky.  To make weights referred to the calibration source we pre and post-multiply both covariance measurements by $\bar{\mathbf{D}}^{-1}$ and $\mathbf{D}^{-1}$ respectively before using them to calculate the beamformer weights. 

This calibration makes the PAF weights independent of drifting electronic gains and independent of electronic path lengths, component bandpasses, and digital sampling synchronisation, all of which vary between the ports of a given PAF and between PAFs on different antennas.  Calibrated weights are more easily and sensibly compared between PAFs on different antennas and with electromagnetic simulations and antenna-range pattern measurements.  Calibrating the electronic path-length and gain variations also results in weights that are predominately real valued and that vary smoothly as a function of each element's position in the focal plane.  This makes it practical to estimate missing weights, for ports corrupted by radio frequency interference or system malfunctions,  by interpolating the calibrated weights of adjacent ports.

\subsection{Results}
\begin{figure}
\centering
\includegraphics[width=0.9\columnwidth,trim=0 0 4mm 8mm, clip]{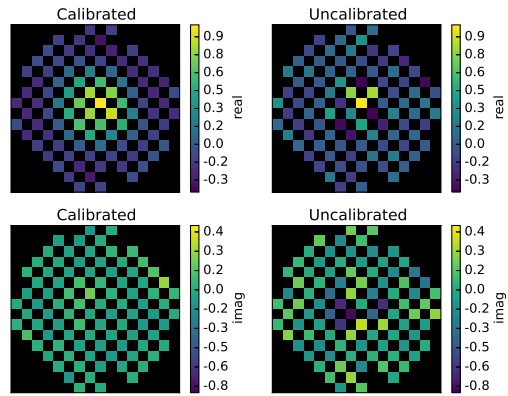}
\caption{Comparing calibrated (left) and uncalibrated (right) weights for a Y-polarisation maximum sensitivity PAF beam at 835~MHz.}
\label{fig_port_cal}
\end{figure}

We have applied the above calibration to new covariance measurements made with the 188-element prototype Mk.~II ASKAP PAF \cite{Hampson2012, Chippendale2015} installed on ASKAP antenna 29.  The new measurements for this work are a repeat of the experiment in \cite{Chippendale2015}.  However, the newly installed prototype of the ASKAP on-reflector calibration system on antenna 29 allowed us to calibrate the covariance data as described above and calculate PAF weights that are calibrated with respect to the on-reflector source.  Fig. \ref{fig_port_cal} shows the Y-polarisation PAF weights at 835~MHz with and without the calibration.  The ``airy pattern'' of the focal field of the reflector can only be seen in the PAF weights after calibration.  The calibration also results in weights that are predominantly real valued, suggesting that it  compensates for the phase differences between ports.

\begin{figure}
\centering
\includegraphics[width=0.9\columnwidth,trim=6mm 3mm 6mm 13mm, clip]{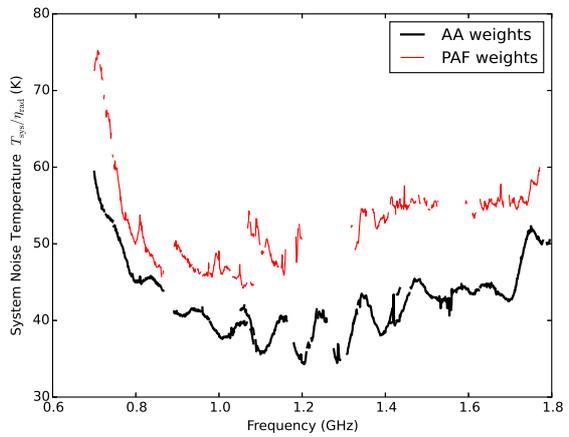}
\caption{System temperature with AA (thick black) and PAF (red) beamforming weights.  PAF weights are formed with the PAF in-reflector observing an astronomical source and transferred to the AA measurement via calibration with an on-reflector source.  The PAF beam is not corrected for absorber illumination efficiency $\alpha$.}
\label{fig_tsys_on_dish}
\end{figure}

Next, we take the AA covariance data measured in \cite{Chippendale2015a} and calibrate it, as described above, with respect to the calibration source suspended over it at the start of each AA Y-factor measurement.  Finally, we beamform the calibrated AA data using both the calibrated PAF weights from the new in-reflector experiment and the AA weights from the original AA experiment in \cite{Chippendale2015a}.  The chequerboard array beam equivalent system noise temperature $T_\text{sys}$ is then calculated via the Y-factor method in \cite{Chippendale2014} and \cite{Chippendale2015a}, but this time for both PAF and AA weights.  Fig. \ref{fig_tsys_on_dish} shows that the resulting AA $T_\text{sys}$ is up to 10~K higher with PAF weights than with AA weights.  However, there are a number of measurement effects we have not yet addressed including the reduced efficiency $\alpha$ with which the broader PAF beam illuminates the hot load in the Y-factor measurement \cite{Chippendale2014} and the potentially higher sidelobes of the PAF beam picking up stray radiation from the Sun and ground. 

Where possible, we carefully controlled factors affecting the AA measurement of the PAF weights.  We configured the digital backend so that the beamformed frequency ranges for the new in-reflector measurements are identical to those used for the AA measurements in \cite{Chippendale2015a}.  The reference radiator installed at the vertex of the reflector for this work is of the same make and model as that used to calibrate the AA measurements in \cite{Chippendale2015a}.  It is carefully installed on the reflector to point directly at the PAF with its polarisation plane at 45$^\circ$ to that of the PAF.  However, the polarisation alignment of the reference radiator with the PAF in \cite{Chippendale2015a} was not ideal, leading to a four-fold suppression of the reference signal in the X-polarisation ports with respect to the Y-polarisation ports and noisier calibrated AA weights for X polarisation.  We therefore only used the Y-polarisation weights in Figs. \ref{fig_port_cal} and \ref{fig_tsys_on_dish}.

In this work, we also made two departures from the beam weight calculation in \cite{Chippendale2015}.  First, we calculated beamformer weights for each polarisation independently by calculating the weights separately for two $94\times94$ covariance matrices, each extracted from the full $188\times188$ matrix by selecting the correlations between the 94 ports of only one linear polarisation at a time.  Second, we performed the eigendecomposition required to calculate the weights on $\mathbf{G}_\text{on}-\mathbf{G}_\text{off}$ instead of on $\mathbf{G}_\text{off}^{-1}\mathbf{G}_\text{on}$.  The first step was required to make a PAF beam of well-defined polarisation that could be reproduced for the AA data.  The second step was taken as it more robustly yielded weights of better sensitivity for this data.  Although seemingly equivalent, working on the difference instead of the ratio of covariances yields differing results for our data.  This may be due to the low signal-to-noise for the astronomical source (Taurus A) used for beamforming.  Its signal is half that of the PAF beam equivalent system noise on a 12~m reflector.

\section{Weights from pattern measurements}
\subsection{Technique}
The beamformed farfield radiation pattern of the array in direction $\Omega$ is
\begin{equation}
  \mathbf{E}_\text{beam}(\Omega) = \textstyle{\sum_{j}}\mathbf{E}_{\text{backend},j}(\Omega)w_{\text{PAF},j}
\end{equation}
where $\mathbf{E}_{\text{backend},j}$ is the radiation pattern of backend port $j$ and $w_{\text{PAF},j}$ is element $j$ of the beamforming weight vector. The backend radiation patterns are obtained from 
\begin{equation}
  \mathbf{E}_{\text{backend},j}(\Omega) = \mathbf{E}_{\text{LNA},j}(\Omega)b_j
\end{equation}
where $\mathbf{E}_{\text{LNA},j}$ is the measured radiation pattern at LNA port $j$ corresponding to backend port $j$.  The coefficient is
\begin{equation}
b_j = {s_{\text{AA},j,\text{backend}}}/{s_{\text{AA},j,\text{LNA}}}
\end{equation}
where $s_{\text{AA},j,\text{backend}}$ is the measured received signal at backend port $j$ when the array is illuminated by the transmitting antenna in the AA noise testing, and $s_{\text{AA},j,\text{LNA}}$ is the corresponding signal at the corresponding LNA port, as computed from the measured LNA-port radiation patterns. 

An estimate of the beamformed receiver  noise contribution  is obtained using the Y-factor method. Thus
\begin{equation}
(T_\text{loss} + T_\text{rec})/{\eta_\text{rad}} = (T_\text{eh}-YT_\text{ec})/(Y-1)
\end{equation}
where $Y=P_\text{h}/P_\text{c}$ is the ratio of measured beamformed powers $P_\text{h}$ and $P_\text{c}$ with hot and cold loads respectively and $T_\text{eh}$ and $T_\text{ec}$ are the corresponding external noise temperature contributions which we compute from
\begin{equation}
  T_\text{ec} = \frac{\int d\Omega T_{b}(\Omega)|\mathbf{E}_\text{beam}(\Omega) |^2}{\int d\Omega|\mathbf{E}_\text{beam}(\Omega)|^2}, \ \text{and}
\end{equation}
\begin{equation}
\begin{split}
  T_\text{eh} &= \frac{\int d\Omega T_{b}(\Omega)|\mathbf{E}_\text{blockedbeam}(\Omega) |^2}{\int d\Omega|\mathbf{E}_\text{beam}(\Omega)|^2} \\
  &+ T_0\frac{\int d\Omega \left(|\mathbf{E}_\text{beam}(\Omega) |^2-|\mathbf{E}_\text{blockedbeam}(\Omega) |^2\right)}{\int d\Omega|\mathbf{E}_\text{beam}(\Omega)|^2}
  \end{split}
\end{equation}
where $T_{b}(\Omega)$ is the brightness temperature distribution of the sky and ground, $T_0$ is the physical temperature of the hot load and $\mathbf{E}_\text{blockedbeam}(\Omega)$ is the beamformed radiation pattern of the array when blocked by the absorber of the hot load. We compute $\mathbf{E}_\text{blockedbeam}(\Omega)$ from the corresponding unblocked radiation pattern $\mathbf{E}_\text{beam}(\Omega)$ by transformation to the nearfield plane of the absorber and then transforming the unblocked field in this plane back to the farfield.

\subsection{Results}
\begin{figure}
\centering
\includegraphics[width=0.8\columnwidth]{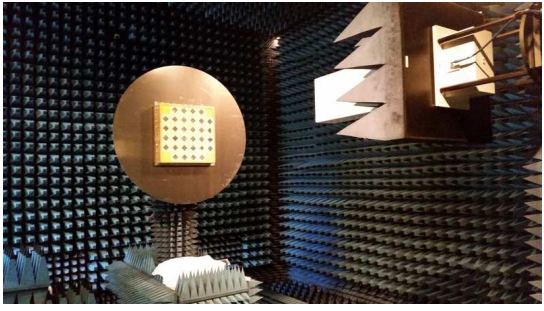}
\caption{40-element PAF in nearfield-measurement range.}
\label{fig_paf_range}
\end{figure}

This approach has been applied to the 40-element prototype of the ASKAP Mk.~II PAF shown in Fig. \ref{fig_paf_range}. The AA noise temperature testing was done at the Parkes Testbed Facility following the approach in \cite{Chippendale2014}.  Array radiation patterns at all LNA ports were measured in a spherical-scanning nearfield range at 0.7~GHz, 0.75~GHz and 0.8--1.8~GHz in 0.1~GHz steps.  These patterns were interpolated to the larger set of frequencies used in the noise-temperature testing. Frequencies corrupted by radio interference are omitted from our results. 

\begin{figure}
\centering
\includegraphics[width=0.8\columnwidth,trim=0 0mm 0 0mm, clip]{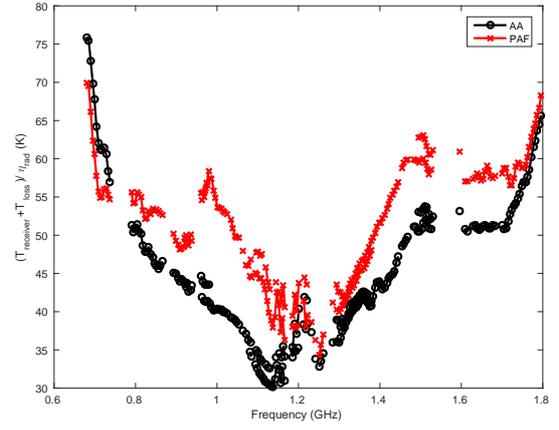}
\caption{Receiver temperature with AA (black) and PAF (red) beamforming weights. AA results are as measured with the transmitting antenna polarised at 45$^\circ$ from horizontal in Fig. \ref{fig_paf_range}. PAF results are for the same polarisation with beamforming conversion derived from array radiation-pattern measurements.}
\label{fig_paf_rx_temp}
\end{figure}

\begin{figure}
\centering
\includegraphics[width=0.8\columnwidth]{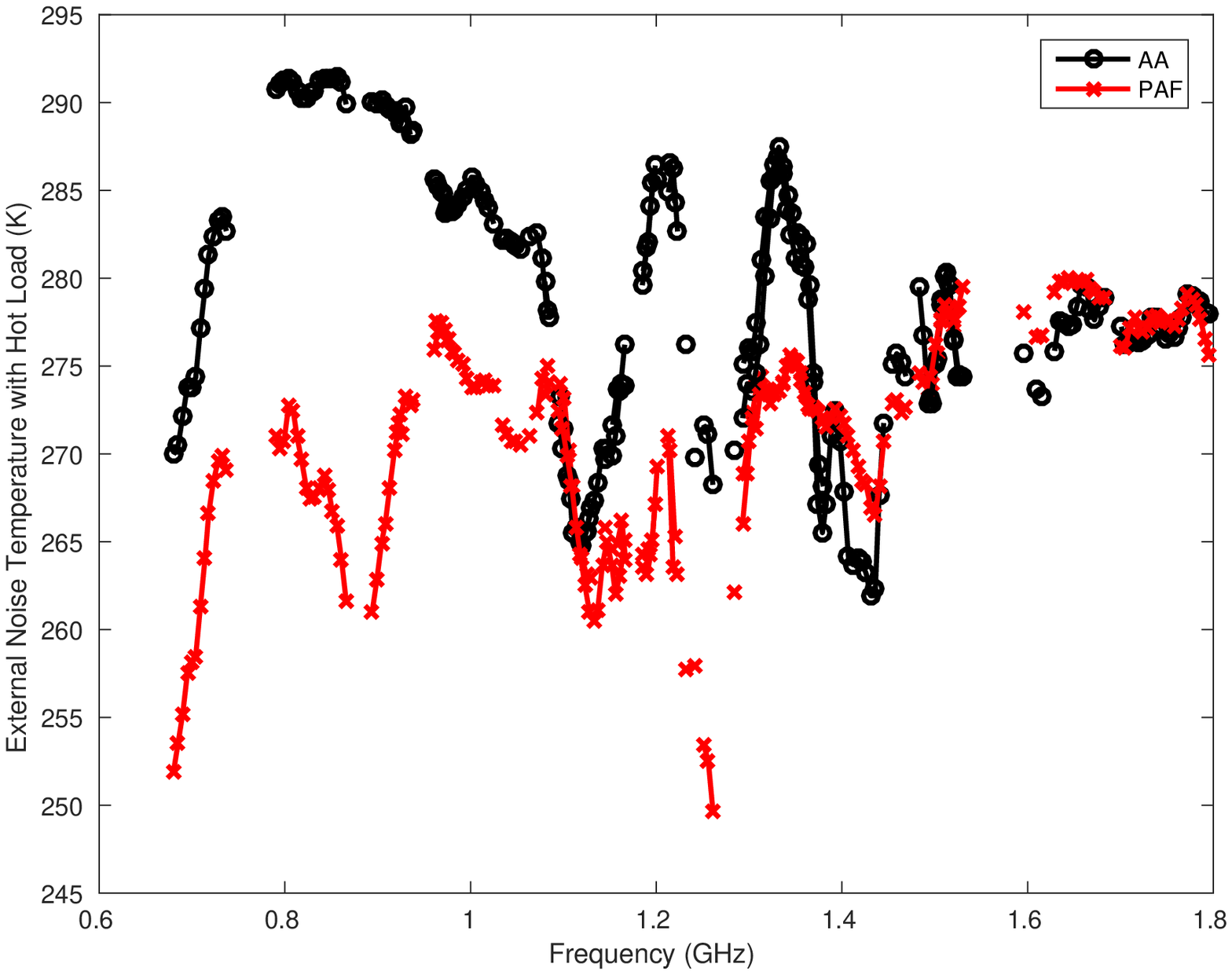}
\caption{Hot-load external temperature with AA (black) and PAF (red) beamforming weights. AA results are measured with the transmitting antenna polarised at 45$^\circ$ from horizontal in Fig. \ref{fig_paf_range}. PAF results are for the same polarisation with beamforming conversion via radiation-pattern measurements.}
\label{fig_load_ext_temp}
\end{figure}

\begin{figure}
\centering
\includegraphics[width=0.85\columnwidth]{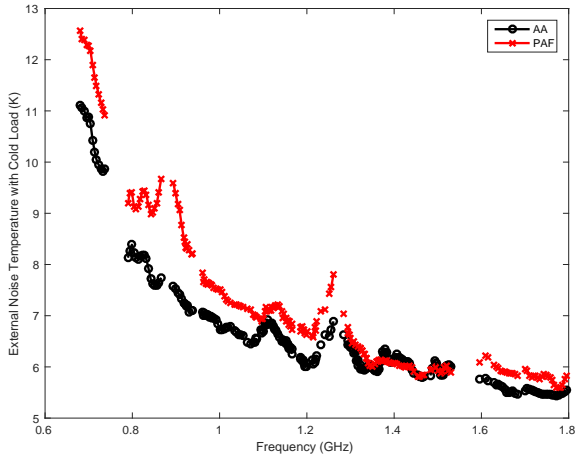}
\caption{Cold-load external temperature with AA (black) and PAF (red) beamforming weights. AA results are measured with the transmitting antenna polarised at 45$^\circ$ from horizontal in Fig. \ref{fig_paf_range}. PAF results are for the same polarisation with beamforming conversion via radiation-pattern measurements.}
\label{fig_cold_ext_temp}
\end{figure}

Figs. \ref{fig_paf_rx_temp} to \ref{fig_cold_ext_temp} compare the receiver, hot-load and cold-load noise contributions for the AA and PAF beamforming weights. The AA polarisation is as measured at 45$^\circ$ from the horizontal in Fig. \ref{fig_paf_range}. The PAF results are for the same polarisation, with the signal conversion derived from the array radiation-pattern measurements applied in computing the beamforming weights. The ASKAP reflector is a 12~m diameter paraboloid with a focal-length to diameter ratio of 0.5. The signal conversion accounts for the finite, 2.018~m distance of the transmitting antenna from the array in the noise-temperature test setup.  The signal conversion factors are computed via \eqref{eq:convfactors} where ${s}_{\text{AA},j}$ is the nearfield radiation pattern of the array at the location and polarization of the transmitting antenna, and ${s}_{\text{PAF},j}$ is the farfield radiation pattern of the combined array and reflector system. Both radiation patterns are computed from measured radiation patterns of the array. The patterns are measured on a spherical surface in the nearfield and transformed to the farfield and back to nearfield planes using well-known spherical-wave expansion and Fourier-transform techniques. 

These results include a significant backend noise contribution from the ASKAP Mk I signal conversion system \cite{Schinckel2011} at the Parkes Test Facility and its non-standard configuration for this work. It is evident from the results changing with different settings of the local oscillator in the frequency conversion system. Further work is required to understand this and the accuracy of this approach to beamforming weight conversion. 

\begin{figure}
\centering
\includegraphics[width=0.85\columnwidth]{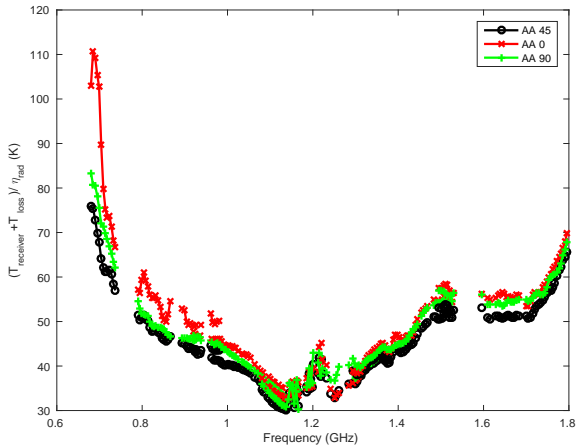}
\caption{Receiver temperature with AA weights for polarisation of 45$^\circ$ (black), 0$^\circ$ (red) and 90$^\circ$ (green) from horizontal in Fig. \ref{fig_paf_range}. The AA 45 results are as measured with the transmitting antenna polarised at 45$^\circ$ from horizontal in Fig. \ref{fig_paf_range}. The other AA results are for the other polarisations with beamforming conversion via radiation-pattern measurements.}
\label{fig_rx_temp_pol}
\end{figure}

Measuring the AA noise temperature with different transmitting antenna polarisations may yield further insight. Fig. \ref{fig_rx_temp_pol} shows the beamforming conversion results for conversion from the 45$^\circ$ polarisation to horizontal and vertical polarisations. The signal conversion factors were again obtained from the measured radiation patterns of the array.  This degree of polarisation dependence is not expected and at this stage the cause is unknown. Initial estimates of the back-end noise contribution also have a significant polarisation dependence.

\section{Conclusion}
Our measurements indicate that the noise temperature of boresight beams from CSIRO's Mk.~II chequerboard PAFs are up to 10~K higher with PAF weights than with AA weights.  This is supported by two independent methods of calculating the in-reflector weights, one via in-reflector measurements with astronomical and on-reflector sources, and the second via antenna-range pattern measurements and reflector modelling.  Determining the precise value of the noise temperature increase will require further development of our techniques.



\section*{Acknowledgment}
A. Chippendale thanks A. Dunning for helpful discussions on calibrating in-reflector PAF weights for application in AA measurements.  M. Leach and R. Beresford designed and installed the on-reflector calibration hardware.  The Australian SKA Pathfinder is part of the Australia Telescope National Facility which is funded by the Commonwealth of Australia for operation as a National Facility managed by CSIRO. This scientific work uses data obtained from the Murchison Radio-astronomy Observatory (MRO), which is jointly funded by the Commonwealth Government of Australia and State Government of Western Australia. The MRO is managed by the CSIRO, who also provide operational support to ASKAP. We acknowledge the Wajarri Yamatji people as the traditional owners of the Observatory site. 



\bibliographystyle{IEEEtran}
\bibliography{IEEEabrv,apj-jour,eucap2016_askap}
%
%
%
%

\end{document}